\documentstyle[psfig,twocolumn]{mn}
\newif\ifAMStwofonts

\newcommand{\etal}{et al. }

\newcommand{\xmm}{{\it XMM-Newton} }

\newcommand{\afe}{{$A_{\rm Fe}$} }
\newcommand{\afep}{{$A_{\rm Fe}$}}

\newcommand{\feka}{{Fe~K$\alpha$} }

\newcommand{\fekalfa}{{Fe~K$\alpha$} }

\newcommand{\fekaeffic}{{$x_{\rm Fe \ K\alpha}$} }
\newcommand{\fekaefficp}{{$x_{\rm Fe \ K\alpha}$}}

\newcommand{\fekb}{{Fe~K$\beta$} }

\newcommand{\fekbeta}{{Fe~K$\beta$} }
\newcommand{\fekbetap}{{Fe~K$\beta$}}

\newcommand{\nh}{$N_{\rm H}$ }
\newcommand{\nhp}{$N_{\rm H}$}
\newcommand{\meannh}{{$\bar{N}_{H}$} }
\newcommand{\meannhp}{{$\bar{N}_{H}$}}

\newcommand{\dovnamep}{Dov\v{c}iak, M.}

\newcommand{\thetaobs}{{$\theta_{\rm obs}$} }
\newcommand{\thetaobsp}{{$\theta_{\rm obs}$}}

\newcommand{\figefficvsnh}{{Fig.~2} }
\newcommand{\figefficvsnhp}{{Fig.~2}}
\newcommand{\figfekcontrib}{{Fig.~1} }
\newcommand{\figfekcontribp}{{Fig.~1}}

\title{On the efficiency of production of the \fekalfa emission line
in neutral matter}

\author[T. Yaqoob, K. D. Murphy, L. Miller, \& T. J. Turner]
{T. Yaqoob$^{1}$, K. D. Murphy$^{2}$, L. Miller$^{3}$ and T. J. Turner$^{4}$ \\
$^{1}$Department of Physics and Astronomy, Johns Hopkins University, Baltimore, MD 21218. \\
$^{2}$MIT Kavli Institute for Astrophysics and Space Research, 77 Massachusetts
        Avenue, NE80-6013, Cambridge, MA 02139. \\
$^{3}$ Department of Physics, University of Oxford, Denys Wilkinson
 Building, Keble Road, Oxford, OX1 3RH, UK. \\
$^{4}$ Department of Physics, University of Maryland Baltimore County,
        1000 Hilltop Circle, Baltimore, MD 21250, USA.
}

\date{Accepted. Received; in original form}

\begin{document}

\maketitle



\begin{abstract} 

The absolute luminosity of the \fekalfa emission line 
from matter illuminated by X-rays 
in astrophysical sources is nontrivial to calculate except
when the line-emitting medium is optically-thin to absorption
and scattering. We characterize the \fekalfa line flux
using a dimensionless 
efficiency, defined as the fraction of continuum photons 
above the Fe~K shell absorption edge threshold energy
that appear in the line. The optically-thin approximation
begins to break down even for column densities as small
as $2 \times 10^{22} \ \rm cm^{-2}$.
We show how to obtain reliable estimates 
of the \fekalfa line efficiency in the case of cold, neutral
matter, even for the Compton-thick regime. 
We find that, regardless of geometry and covering factor,
the largest \fekalfa line efficiency is attained
well before the medium becomes Compton-thick.   
For cosmic elemental abundances it is difficult
to achieve an efficiency higher than a few percent under the
most favorable conditions and lines of sight. For a given
geometry, Compton-thick lines-of-sight may have 
\fekalfa line efficiencies that are orders of magnitude less
than the maximum possible for that geometry. Configurations
that allow unobscured views of a Compton-thick
reflecting surface are capable of yielding the highest
efficiencies. 
Our results can be used to estimate the predicted flux
of the narrow \fekalfa line at $\sim 6.4$~keV
from absorption models in AGN. In particular we show that
contrary to a recent claim in the literature,
absorption-dominated
models for the relativistic \fekalfa emission line in MCG~$-$6-30-15
{\it do not} over-predict the narrow \fekalfa line
for any column density or covering factor.

\end{abstract}



\section{Introduction}
\label{torusintro}

The narrow
(FWHM~$<10^{4} \ \rm km \ s^{-1}$) fluorescent
\fekalfa emission line at $\sim 6.4$~keV
is a ubiquitous feature in the X-ray spectra of
both type~1 and type~2 active galactic nuclei (AGNs),
its centroid energy indicating an origin in cold, neutral matter
(e.g. Sulentic \etal 1998;
Yaqoob \& Padmanabhan 2004;
Levenson \etal 2006, and references therein).
The \fekalfa line at $\sim 6.4$~keV is also an important diagnostic in
some X-ray binary systems (e.g. White \etal 1995;
Watanabe \etal 2003;
Paul \etal 2005; Miller 2007, and references therein).
The equivalent width (EW) and luminosity of the \fekalfa line are nontrivial
to calculate, except when the total Fe~K band ($\sim 6$--$7$~keV) optical depth 
of the line to absorption and
scattering is $\ll 1$. In general, 
the observed EW and line flux depend,
at the very least, on the geometry, column density
of the line-emitting matter, covering factor, element abundances,
and the orientation of the structure relative to the observer's line of sight.
Using cosmic abundances from Anders \& Grevesse (1989) and photoelectric
cross-section from Verner \etal (1996), it is straightforward to show
that the optical depth of the \fekalfa line to absorption plus 
Compton scattering
is $\tau_{\rm tot} \sim 0.027N_{22}$, where $N_{22}$ is the column density, $N_{H}$, in units
of $10^{22} \rm \ cm^{-2}$. Thus, $\tau_{\rm tot}$ is already $\sim 0.1$ for
column densities as low as $4 \times 10^{22} \rm \ cm^{-2}$ so that we
expect EWs and line fluxes calculated using an optically-thin approximation
to break down even for column densities lower than $\sim  10^{23} \rm \ cm^{-2}$.
Indeed, using detailed Monte Carlo simulations of a toroidal X-ray reprocessor,
Murphy \& Yaqoob (2009; hereafter MY09) showed  that geometrical and inclination-angle effects
begin to become important for $N_{H} \sim 10^{23} \rm \ cm^{-2}$.

Whilst detailed calculations of the \fekalfa line EW for various geometries
are abundant in the literature (e.g. 
Leahy \& Creighton 1993; Ghisellini \etal 1994; 
 Krolik \etal 1994; Nandra \& George 1994; 
Matt \etal 1999; Ikeda, Awaki, \& Terashima 2009;
MY09), detailed information on the \fekalfa line {\it flux}
(or luminosity) is lacking.
In some situations, one may be interested in the absolute flux of the
\fekalfa line (in addition to, or instead of the EW). For example,
we may want to know the fraction of the number of photons 
(or fraction of the luminosity)
in the continuum 
above the Fe~K photoelectric absorption edge threshold energy 
that appear in
the \fekalfa emission line. Since the reprocessed, scattered X-ray continuum
at any particular energy is a complicated function of the column density,
geometry, and system inclination angle, the \fekalfa emission-line
flux cannot be trivially calculated from the EW. 
In some situations, it may be more appropriate, or easier, 
to compare and interpret the \fekalfa line {\it flux} from different models 
(as opposed to the EW). In the present paper we calculate the
{\it efficiency} of production of the \fekalfa line, expressed
as the ratio of the number of photons in the emission line to the
number of photons in the illuminating continuum, above
the Fe~K edge absorption threshold energy. We utilize Monte Carlo
results from the toroidal X-ray reprocessor model of
Murphy \& Yaqoob, described 
in detail in MY09, but we also discuss the 
generalization of our results to other geometries.

The paper is organized as follows.
In \S\ref{mytorusmodel}
we give an overview of the details of the Monte Carlo model.  
In \S\ref{lineflux} we 
present the results for calculations of the efficiency of production
of the \fekalfa line in the context of a toroidal X-ray reprocessor model.
In \S\ref{othergeom} we discuss
the generalization of our results for other geometries and
arbitrary covering factors.
In \S\ref{application} we apply our results to the well-known
Seyfert galaxy MCG~$-$6-30-15.
We summarize our results and conclusions in \S\ref{torusconcl}.

\section{Toroidal X-ray reprocessor model}
\label{mytorusmodel}

We have constructed a Monte-Carlo code to calculate grids of Green's 
functions to model the passage of X-rays through a toroidal reprocessor. 
The model, and some basic results, have
been described in detail in MY09. Here we give a brief overview of the critical
assumptions that the model is based upon.
Our geometry is an azimuthally-symmetric 
doughnut-like torus with a circular cross-section,
characterized by only two parameters, namely
the half-opening angle, $\theta_{0}$, and the
equatorial column density, \nh (see MY09 for details). 
The inclination angle between the observer's line of sight and the 
symmetry axis of the torus is given by
\thetaobsp, where \thetaobs$=0^{\circ}$ corresponds to a face-on 
observing angle and \thetaobs$=90^{\circ}$
corresponds to an edge-on observing angle (see MY09 for exact
definitions of the face-on and edge-on angle bins).
We assume that the X-ray source emits isotropically and that 
the reprocessing material is uniform and
essentially neutral (cold). 
For illumination by an X-ray source that is emitting isotropically,
the mean column density, integrated over all incident
angles of rays through 
the torus, is \meannhp~$=(\pi/4)$\nhp. 
The column density may also be expressed in terms of the Thomson depth: 
$\tau_{\rm T} = (11/9) N_{\rm
H}\sigma_{\rm T} \sim 0.81N_{24}$ where $N_{24}$ is the column density in 
units of $10^{24} \rm \
cm^{-2}$. 
Here, we have employed
the mean number of electrons per H atom, $\frac{1}{2}(1+\mu)$, 
where $\mu$ is the mean molecular weight.
Note that this assumes that the abundance of He is 10\% by 
number and that the number of
electrons from all other elements aside from H and He is  
negligible (i.e. $\mu = 13/9$).

The value of $\theta_{0}$ for which we have calculated
a comprehensive set of Green's functions is $60^{\circ}$,
for $N_{H}$ in the range $10^{22-25} \ \rm cm^{-2}$, using input energies
up to 500~keV-- see MY09 for details. For
$\theta_{0}=60^{\circ}$, the solid angle subtended by the torus at the
X-ray source, $\Delta\Omega$, is $2\pi$ 
(and we refer to $[\Delta\Omega/(4\pi)]$ as a covering factor, 
which in this case is 0.5). 
Our model employs a
full relativistic treatment of Compton scattering,
using the full differential and total Klein-Nishina Compton-scattering
cross-sections. 
We utilized photoelectric absorption cross-sections for 30 elements as 
described in Verner \& Yakovlev (1995) and
Verner \etal (1996).  
We used Anders and Grevesse (1989) elemental cosmic abundances in
our calculations.
For the \fekalfa and \fekbeta fluorescent emission lines,
we used rest-frame line energies of 6.400 and 7.058~keV
respectively, appropriate for neutral matter (e.g.
see Palmeri \etal 2003). 
We used a fluorescence yield for Fe of 0.347 
(see Bambynek \etal 1972) and an \fekb to \feka 
branching ratio of 0.135 
(representative of the range of 
experimental and theoretical values discussed in Palmeri \etal 2003,
see also Kallman \etal 2004).

\section{Efficiency of production of the \feka line }
\label{lineflux}

Observational measurements of the 
equivalent width (EW) and/or flux of the fluorescent \feka emission line
in the literature generally refer to the zeroth-order emission only, 
since the line is typically
fitted with a Gaussian.  The scattered flux 
(i.e., the Compton shoulder) can carry up to $\sim 40\%$ of
the zeroth-order flux, depending on the geometry
and column density of the matter (e.g. Matt 2002; MY09).
However, the scattered part of the line, or Compton shoulder,
is generally not fitted (usually because it is not
detected), or, if it is fitted, a separate EW or flux is quoted.  
On the other hand, part of the Compton shoulder may be
blended with the zeroth-order line since most of the flux
of the Compton shoulder spans an energy
range that stretches from the zeroth-order line energy
to $\sim 6.25$~keV. The amount of blending depends on the
column density, geometry, and orientation of the line emitter,
as well as the velocity broadening and instrumental
spectral resolution.
However, in the remainder of the paper, 
the \fekalfa line flux, intensity, luminosity, or EW,
will always refer to the  
zeroth-order (unscattered) part of the
emission line. Corrections for the Compton shoulder should be
made that are appropriate for a particular situation.

We define the efficiency of production of the \fekalfa emission
line, $x_{\rm Fe \ K\alpha}$, as the ratio of the line flux, 
$I_{\rm Fe K\alpha}$, to the integrated
flux in the incident X-ray continuum above the
Fe~K absorption edge threshold energy, $E_{K}$:

\begin{eqnarray}
x_{\rm Fe \ K\alpha} & \equiv & \frac{I_{\rm Fe \ K\alpha}}
{\int_{E_{K}}^{\infty}{ N(E) \ dE}}.
\label{eqn:effgendef}
\end{eqnarray}

Here, $E_{K}$ is the threshold energy for Fe~K-shell absorption,
and we adopt the Verner \etal (1996) value of $7.124$~keV.
Assuming
a power-law incident continuum, with photon index $\Gamma$,

\begin{eqnarray}
x_{\rm Fe \ K\alpha} & = & \frac{I_{\rm Fe \ K\alpha, n}}
{\int_{E_{K}}^{\infty}{ E^{-\Gamma} dE}} \\
& = & I_{\rm Fe \ K\alpha, n} \ E_{K}^{\Gamma-1} \ (\Gamma-1)
\ \ \ \ \ \ \ \ \ \ \ \  (\Gamma > 1),
\label{eqn:efficdef}
\end{eqnarray}

where $I_{\rm Fe \ K\alpha, n}$ refers to the line flux renormalized
to an incident continuum that has a monochromatic flux
of 1~photon $\rm cm^{-2} \ s^{-1} \ keV^{-1}$ at 1 keV. The efficiency
does not of course depend on the absolute normalization of
the continuum. Note that the definition for the efficiency
explicitly assumes that the intrinsic X-ray continuum emission is
isotropic and emitted into a solid angle of $4\pi$.
However, $I_{\rm Fe \ K\alpha, n}$ refers to the {\it observed}
line flux so the efficiency may have an inclination-angle dependence.
In the definition of \fekaeffic we choose
to use an upper limit of infinity because we find from our
Monte Carlo simulations that when the medium becomes Compton-thick,
downscattering of photons with initial energies much higher
than even 20~keV to lower energies enables those high-energy photons
to make significant contributions to the \fekalfa line flux.
This is illustrated in \figfekcontrib which shows, 
for several column densities,
the number of \fekalfa emission-line photons
resulting from continuum photons injected into the medium at an energy 
$E$, as a fraction of
the number of line photons resulting from continuum photons injected
at 7.2~keV (just above the Fe~K absorption edge).
For example, it can be seen that for $N_{H} = 5 \times 10^{24} \ \rm cm^{-2}$
(dotted line, \figfekcontribp), the curve breaks
at $\sim 11$~keV, becoming flatter at higher energies.
For $N_{H} = 10^{25} \ \rm cm^{-2}$ 
(upper solid line, \figfekcontribp) the curve breaks
and becomes flatter at $\sim 15$~keV.
An upper limit of infinity on
the integral in equation~\ref{eqn:effgendef} is also less arbitrary than 
choosing a specific value. In practice, the incident X-ray
continuum in AGN will steepen and cut off at high energies (a few
hundred keV in AGN -- e.g. see Dadina 2008, and references therein).
The effect of such a cut-off will give a somewhat larger
efficiency than that given by equation~\ref{eqn:effgendef}. However,
our purpose is to facilitate the estimation of the absolute \fekalfa
line flux; corrections required for different incident
continua can easily be calculated. It is also worth mentioning
that the absorption opacity above the Fe~K edge is {\it not}
due mostly to Fe. The ratio of the Fe~K shell opacity to the
sum of all of the other absorption opacities (for the cross-sections
and abundances that we have adopted) is 0.51 just above the
Fe~K shell threshold energy, and never exceeds 0.61.

\begin{figure}
\centerline{
\psfig{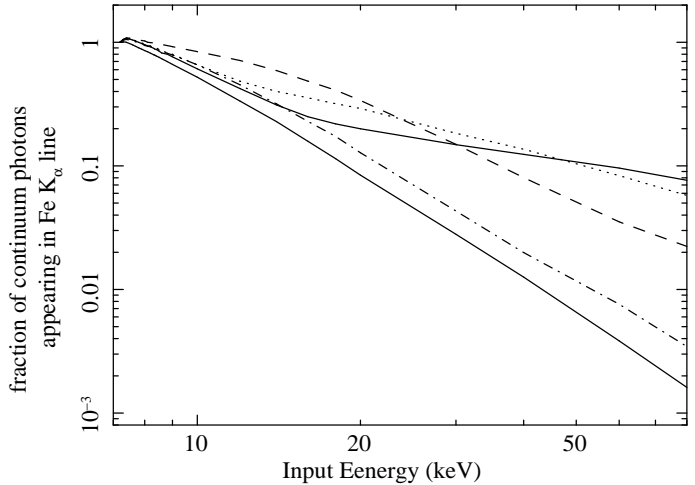}}
\caption{\footnotesize  Monte Carlo results for
the number of escaping \fekalfa emission-line photons
resulting from monoenergetic continuum photons injected into the 
toroidal X-ray reprocessor of Murphy \& Yaqoob (2009) at an energy
$E$, as a fraction of
the number of line photons resulting from continuum photons injected
at 7.2~keV. Results are shown for five equatorial column densities:
$2 \times 10^{23} \ \rm cm^{-2}$ (lower solid curve),
$5 \times 10^{23} \ \rm cm^{-2}$ (dot-dashed curve),
$2 \times 10^{24} \ \rm cm^{-2}$ (dashed curve),
$5 \times 10^{24} \ \rm cm^{-2}$ (dotted curve), and
$10^{25} \ \rm cm^{-2}$ (upper solid curve). For each column density, the
\fekalfa line photons that escape the medium are summed over
all escape directions. As the
reprocessor becomes more and more 
Compton-thick, the relative contribution to the \fekalfa line
from high-energy continuum photons increases significantly.
}
\end{figure}

\begin{figure}
\centerline{
\psfig{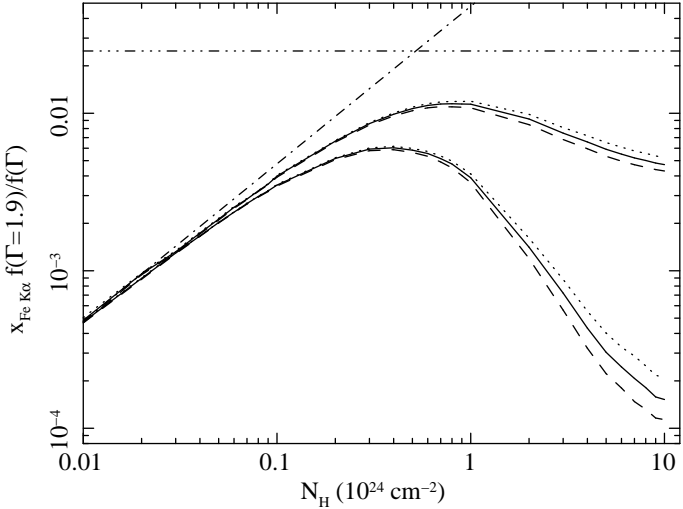}}
\caption{\footnotesize The \feka line efficiency,
\fekaefficp, 
(defined in equation~\ref{eqn:effgendef}--see text) versus 
equatorial column density, $N_{H}$, for a toroidal X-ray
reprocessor model with $[\Delta\Omega/(4\pi)]=0.5$.
When comparing with other geometries, the mean column density,
$(\pi/4)N_{H}$, should be used.
Solid curves correspond to 
a power-law incident continuum with $\Gamma=1.9$.
Dotted curves are for the same geometry but for $\Gamma=1.5$,
multiplied by the factor $f(\Gamma=1.9)/f(\Gamma=1.5)$, where
$f(\Gamma)$ is given by equation~\ref{eqn:fgam}. 
Dashed curves correspond to $\Gamma=2.5$, multiplied by the
factor $f(\Gamma=1.9)/f(\Gamma=2.5)$.
For each pair of curves for a given value of $\Gamma$, the
upper and lower curves correspond to the face-on and edge-on
inclination-angle bins respectively.
The dot-dashed line
shows the relation for the optically-thin limit (for $\Gamma=1.9$),
from equation~\ref{eqn:thinlimit}. The dashed,
triple-dotted (horizontal) line shows the value of
\fekaeffic for a standard, face-on, semi-infinite disk that 
has a Compton-depth $>10$ ($N_{H}> 1.25 \times 10^{25} \rm \ cm^{-2}$). 
}
\end{figure}

In the optically-thin limit 
for which absorption and scattering optical depths in the
Fe~K band ($\sim 6$--$7$~keV) are $\ll 1$, we can obtain an analytic expression
for \fekaefficp. We can adopt the optically-thin expression for
the \fekalfa line EW in MY09 and multiply by the continuum
at 6.4~keV in this limit. We get

\vspace{1cm}

\begin{eqnarray}
x_{\rm Fe \ K\alpha}
& = &
0.00482
\ \left(\frac{\Delta\Omega}{4\pi}\right)
\ \left(\frac{\omega_{K}}{0.347}\right)
\ \left(\frac{\omega_{K\alpha}}{\omega_{K}}\right)
\ \left(\frac{\rm A_{\rm Fe}}{4.68 \times 10^{-5}}\right) \nonumber \\
& \times & \ \left(\frac{\sigma_{0}}{3.37\times10^{-20} \rm \ cm^{2}}\right)
\ N_{22}
\ f(\Gamma)
  \ \ \ \ \   (\Gamma > 1),
\label{eqn:thinlimit}
\end{eqnarray}

where

\begin{eqnarray}
f(\Gamma) 
& \equiv &
\frac{\Gamma-1}{\Gamma + \alpha -1}
\label{eqn:fgam}
\end{eqnarray}

The quantity $[\Delta\Omega/(4\pi)]$ is the fractional solid angle that 
the line-emitting matter subtends at the
X-ray source (and we refer to it synonymously as the covering factor).  
For the MY09 torus model, 
$N_{22}=\bar{N}_{H}/(10^{22} \rm \ cm^{-2})$,
where $\bar{N}_{H}=(\pi/4)N_{H}$ (see \S\ref{mytorusmodel}).
The K-shell fluorescence yield is given by
$\omega_{K}$, and $\omega_{K\alpha}$ is the yield for the
\fekalfa line only (i.e. not including \fekbetap). Using
our adopted value of 0.135 for the \fekbetap/\fekalfa branching ratio,
 $\omega_{K\alpha}/\omega_{K}=0.881$.
The quantity 
$A_{\rm Fe}$ is the Fe abundance relative to Hydrogen
($4.68 \times 10^{-5}$ is the cosmic value in Anders \& Grevesse 1989).
The quantity $\sigma_{0}$
is the Fe~K shell absorption cross-section at the Fe~K edge energy, $E_{K}$,
and $\alpha$ is the power-law index of
the cross-section as a function of energy. 
For the Verner \etal (1996) data that we have adopted,
$\sigma_{0} = 3.37 \times 10^{-20} \ \rm cm^{2}$,
$\alpha=2.67$ (see MY09).

Using the MY09 toroidal X-ray reprocessor Monte Carlo 
results for $[\Delta\Omega/(4\pi)]=0.5$
and an input power-law continuum with $\Gamma=1.9$, we constructed
curves of the \fekalfa line efficiency,
\fekaefficp, versus \nh using equation~\ref{eqn:effgendef}.
\figefficvsnh shows two such curves (solid lines), the upper and
lower curves corresponding to the face-on and edge-on inclination-angle
bins respectively. It can be seen that both curves show 
similar behavior. As \nh is increased, \fekaeffic first increases
but then turns over, reaching a maximum for \nh somewhere between
$\sim 3-8 \times 10^{23} \ \rm cm^{-2}$, depending on the
inclination angle. This is because the escape of \fekalfa line
photons from the medium after they are created is significantly
impeded for larger column densities
by absorption and scattering opacity that is
relevant at the line energy. For the edge-on angle bin the
maximum \fekalfa line efficiency is attained {\it well before the
medium becomes Compton-thick} 
(i.e. before \nh becomes
as high as $1.25 \times 10^{24} \ \rm cm^{-2}$). 
This is also true of the angle-averaged maximum, which we find
occurs at \nh~$\sim 4 \times 10^{23} \ \rm cm^{-2}$.
The position of the turnover can be understood as approximately
corresponding to a situation for which the average
optical depth to absorption plus scattering for
the zeroth-order \fekalfa line photons is of order
unity. The single-scattering
albedo (ratio of scattering to total cross-section)
at 6.4~keV, for the element abundances that
we have adopted, is $\sim 0.31$. The optical depth to scattering
is $\sim 0.81\bar{N}_{H}/(10^{24} \rm \ cm^{-2})$,
or $\sim 0.81(\pi/4)N_{24}$, so putting $0.81(\pi/4)N_{24}/0.31 \sim 1$ gives
$N_{24} \sim 0.5$ for the peak value of
\fekaefficp, agreeing well with our Monte Carlo
results. For the covering factor of 0.5, it can seen that the
\fekalfa line efficiency is never more than $\sim 1.15\%$.

Also shown in \figefficvsnh (dot-dashed line) is the optically-thin
limit from equation~\ref{eqn:thinlimit}. It can be seen that the 
Monte Carlo curves converge to this optically-thin limit, but only
for column densities $<2 \times 10^{22} \ \rm cm^{-2}$.
It can also be seen that even for column densities as low
as $10^{23} \ \rm cm^{-2}$ (when the Thomson depth is only 0.08),
errors as large as $\sim 40\%$ can be incurred if one uses
the optically-thin limit to calculate \fekalfa line fluxes
and luminosities. This is because the important quantity
that determines the observed fraction of \fekalfa line
photons is the {\it total} opacity due to photoelectric
absorption and scattering. At 6.4~keV, the ratio of the
absorption to scattering cross-section is $\sim 7:3$ in our model.
{\it Our results show that the optically-thin approximation should
never be used to calculate or estimate \fekalfa line fluxes
for \nhp~$> 4 \times 10^{22} \ \rm cm^{-2}$ or so}.
The error incurred in neglecting the opacity for \fekalfa line photons
to escape the medium exceeds an order of magnitude 
(a factor of $\sim 11$) 
for a column density of $2 \times 10^{24} \ \rm cm^{-2}$
for the face-on bin, and is nearly two orders of magnitude for the edge-on bin
(a factor of $\sim 71$).

Although it is well-known that flatter incident continua
give large \fekalfa line EWs and fluxes (e.g. MY09 and references therein),
{\it the \fekalfa line efficiency is smaller for flatter incident continua}
(lower values of $\Gamma$). The reason is that steeper
continua give a greater relative weight to the Fe~K-shell absorption
cross-section where it is largest (i.e. at lower energies down
to the Fe~K edge threshold energy).
This can be seen clearly in equation~\ref{eqn:thinlimit} for the
optically-thin limit, but it is true in general. We have calculated
\fekaeffic using various values of $\Gamma$ from the MY09 toroidal
X-ray reprocessor Monte Carlo results and find, rather 
surprisingly, that the \fekaeffic versus \nh curves for 
different power-law slopes can be accounted for by the
optically-thin factor, $f(\Gamma)$ in equation~\ref{eqn:fgam}, even up
to column densities of $\sim 2 \times 10^{24} \ \rm cm^{-2}$.
In other words if we have curves of \fekaeffic versus \nh for
$\Gamma=1.9$, we can estimate the relation for a different
value of $\Gamma$ by multiplying the \fekaeffic versus \nh
relation for $\Gamma=1.9$ by the factor $f(\Gamma)/f(\Gamma=1.9)$.
This is illustrated in \figefficvsnh, where
we have applied the process in reverse to the Monte Carlo
results for the torus for $\Gamma=1.5$ (dotted curves) and
$\Gamma=2.5$ (dashed curves), for the face-on and edge-on angle
bins. This range in $\Gamma$ of 1.5--2.5 covers
values of the X-ray power-law continuum slope
that is observationally-relevant for AGN.
That is, we multiplied the actual Monte Carlo results for
$\Gamma=1.5$ and $\Gamma=2.5$ by the
factor $f(\Gamma=1.9)/f(\Gamma=1.5)$ and $f(\Gamma=1.9)/f(\Gamma=2.5)$
respectively. In the case
that the approximation were perfect,
the dotted, dashed, and solid curves in \figefficvsnh would be identical
for a given orientation of the reprocessor. We see
from \figefficvsnh that the error incurred in using this 
approximation to estimate \fekaeffic from  
the $\Gamma=1.9$ results is {\it in the worst case} only $\sim 11\%$
for the face-on angle bin (i.e. for the largest column density
of $10^{25} \ \rm cm^{-2}$ and
$\Gamma=2.5$). For the edge-on angle bin the
approximation is good to  $\sim 18\%$ or better up to
\nhp~$\sim 2 \times 10^{24} \ \rm cm^{-2}$ but increases with
\nh up to $\sim 37\%$ for \nhp~$\sim 10^{25} \ \rm cm^{-2}$
and $\Gamma=2.5$.
We note that $f(\Gamma=1.5)/f(\Gamma=1.9) \sim 0.63$, and
$f(\Gamma=2.5)/f(\Gamma=1.9) \sim 1.43$, so the absolute
\fekalfa line efficiencies have a factor $\sim 2.3$
range for a given column density as $\Gamma$ varies in the
observationally relevant range of 1.5--2.5. 
In the remainder of the present paper we shall refer to
values of \fekaeffic for $\Gamma =1.9$. Results corresponding
to other values of $\Gamma$ can then be estimated using
the method described above. 

For reference, \figefficvsnh also shows (dashed, triple-dotted line)
the efficiency of
\fekalfa line production for the standard case of a face-on,
centrally-illuminated semi-infinite,
Compton-thick disk (e.g. George \& Fabian 1991)
that subtends the same solid angle at the X-ray source as our
default torus model (i.e. $2\pi$). We calculated \fekaeffic
for the case of the disk using the {\tt hrefl} disk-reflection
continuum model (see \dovnamep, Karas, \& Yaqoob 2004) for
a power-law incident continuum with $\Gamma=1.9$, and an EW
for the \fekalfa line of 143~eV (from George \& Fabian 1991).
The \fekalfa line efficiency for the disk is $\sim 2.5\%$, and
is a constant in \figefficvsnh because it corresponds to
a disk with a Compton depth $>10$ ($N_{H} >1.25 \times 10^{25} 
\ \rm cm^{-2}$).
The reason why 
the disk has a higher value of \fekaeffic than the maximum
for the torus (obtained for face-on reflection), 
even though both subtend a solid angle of $2\pi$ at the
X-ray source, is due to geometrical effects that have
been discussed in detail in MY09. 

\subsection{Effect of Fe abundance and mild ionization}

A higher abundance of Fe, $A_{\rm Fe}$, will increase \fekaefficp.
In the optically-thin limit the increase in \fekaeffic will be in
direct proportion to \afep. However, the total absorption
opacity at 6.4~keV also increases. If \afe increases by a factor
10, the absorption opacity at 6.4~keV increases by a factor of 2.5
and the total opacity to absorption plus scattering increases
by a factor 2.0. Therefore, as the column density increases,
the value of \fekaeffic as a function of \nh will reach a maximum
for smaller values of \nh relative to the case for cosmic Fe 
abundance. The maximum value for
\fekaeffic must be much less than the optically-thin value
for these values of \nh and \afe (equation~\ref{eqn:thinlimit}).
Therefore, even if we increase $A_{\rm Fe}$ by a factor of
10 relative to the cosmic value, 
the maximum \fekalfa line efficiency will only increase by
much less than a factor of 10. We performed some
Monte Carlo calculations using the MY09 toroidal reprocessor model
with the Fe abundance increased by a factor 10 relative
to the cosmic value and found that the maximum
\fekalfa line efficiency increased by a factor of $\sim 2.8$,
to $\sim 3.2\%$, attained for a face-on inclination angle,
and \nhp~$\sim 3-4 \times 10^{23} \ \rm cm^{-2}$.
For an edge-on inclination angle the maximum value
of \fekaeffic increased by a factor of $\sim 3.3$ relative to
the cosmic Fe abundance value, to $\sim 2\%$, for 
\nhp~$\sim 1-2 \times 10^{23} \ \rm cm^{-2}$.

The calculations in the present paper pertain specifically
to neutral matter.
Ionization states of Fe higher than Fe~{\sc xvii} are not
relevant to our discussion because we are making predictions
for the \fekalfa line in AGN that peaks at 6.4~keV. 
Even ionization states of Fe~{\sc xvii} give an \fekalfa line
energy that is higher than observed ($\sim 6.43$~keV --e.g.
Palmeri \etal 2003; Mendoza \etal 2004),
and ionization states of Fe~{\sc xii} or lower are likely
to be more appropriate (e.g. see Yaqoob \etal 2007).
Mild ionization will have the effect of increasing the
\fekalfa line efficiency due to the reduced absorption
opacities impeding the escape of line photons.
We have investigated the effect of mild ionization on the
\fekalfa line efficiency in the context of the MY09 toroidal
reprocessor model using a simple approach that gives a
conservative upper limit on the \fekalfa line efficiency.
We used the {\sc xstar} photoionization code (e.g. Kallman \etal 2004)
to estimate the ionization parameter, $\xi$
\footnote{Here we use the definition $\xi \equiv L_{\rm ion}/(n
r^{2})$, where $L_{\rm ion}$ is the ionizing luminosity 
between 1--1000 Rydberg, $n$ is the proton density,
and $r$ is the distance between the ionizing source
and the illuminated surface of the matter.}, for which the
Fe~K edge energy moves to $\sim 7.4$~keV (appropriate
for Fe~{\sc xii}--see Verner \etal 1996), 
up from the neutral Fe value of $\sim 7.1$~keV.
Although the geometry assumed by {\sc xstar} is a spherical
shell, our simple approach will still give a useful upper limit
because in practice only the inner surface of the toroidal
reprocessor will be ionized but we simply substituted the
total absorption opacities for all elements 
throughout the torus with the reduced opacities calculated by
{\sc xstar}. We used a column density
of $5 \times 10^{23} \rm \ cm^{-2}$ but calculated
the absorption cross-sections per unit column density
so that we could scale the cross-section
for other column densities when applied in the
toroidal Monte Carlo code.
For a column density of $5 \times 10^{23} \rm \ cm^{-2}$
we found an upper limit on $\log{\xi}$ of 1.8
for the Fe K edge threshold energy to remain
below $\sim 7.4$~keV. Thus we used the 
total absorption cross-section for $\log{\xi}=1.8$
throughout the torus. Again, this procedure will
underestimate the absorption at the energy
of the \fekalfa line,
consistent with the goal of obtaining a conservative upper limit
to the \fekalfa line efficiency. We retained the 
K-shell cross-section for neutral Fe because it decreases
slowly with increasing ionization state. We also
retained the fluorescence yield for neutral Fe, which increases
slowly with increasing ionization state. The changes in
both K-shell cross-section and fluorescence yield are less than
15\% up to Fe~{\sc xvii}, and the opposite sense
of the changes tend to compensate each other (e.g. Bambynek \etal 1972)
to some extent.
We obtained the result that, using these very conservative
assumptions, the maximum \fekalfa line efficiency
increases to 1.8\%, the maximum being obtained for
the same column density and orientation as the strictly
neutral torus (i.e.
a face-on inclination angle and $N_{H} \sim 7-8 \times
10^{23} \ \rm cm^{-2}$, or $\bar{N}_{H} \sim 5.5-6.2 \times
10^{23} \ \rm cm^{-2}$). This corresponds to a maximum
enhancement in the \fekalfa line efficiency of $\sim 50\%$
for mild ionization compared to the case of purely neutral matter.

\section{Other geometries and covering factors}
\label{othergeom}

In the limit when a
medium is optically-thin to absorption and scattering in the
Fe~K band ($\sim 6$-$7$~keV), the relation between the
\fekalfa line efficiency and \nh is identical
for all geometries for a given
covering factor. The efficiency, \fekaefficp, can then
be found from equation~\ref{eqn:thinlimit} for any covering factor.
The column density that is relevant for a given geometry
is the average over all incident rays from the X-ray source,
over all lines-of-sight 
that intercept the reprocessing structure.
In terms of achieving the highest \fekalfa line efficiency,
there is a trade-off with the solid angle subtended
by the structure at the X-ray source, or covering factor.
Larger solid angles obviously intercept more continuum
photons to produce
more \fekalfa line photons, but if the solid angle becomes
too large, the escape of \fekalfa line photons could be impeded
(e.g. see discussion in Ikeda \etal 2009). The optimum
opening angle for the maximum \fekalfa line efficiency
depends on the geometry and the observer's line-of-sight
to the reprocessor. The maximum \fekalfa line efficiency
will always be achieved for lines of sight that are
unobscured. 
For the toroidal geometry of MY09, a
maximum efficiency, $\sim 2.8\%$, is achieved 
for a covering factor of $[\Delta\Omega/(4\pi)]\sim 0.8-0.9$
for $N_{H} \sim 8 \times 10^{23} \ \rm
cm^{-2}$  (and a face-on orientation), before
the medium becomes Compton-thick. 

The MY09 Monte Carlo code is adaptable for different geometries
and we have used it to compare the \fekalfa line efficiency
obtained from a toroidal geometry with the case of
a centrally-illuminated, fully-covering sphere, and an
externally-illuminated sphere.
For the former, 
we found that \fekaeffic has a 
broad peak as a function of (radial) $N_{H}$, in the
region $\sim 3-5 \times 10^{23}
\rm \ cm^{-2}$, achieving a maximum value of $\sim 1.6\%$.
In the case of the externally-illuminated sphere, we
assumed a parallel beam incident on one side of the sphere
and calculated \fekaeffic as a function of the angle between
the escaping line photons and the incident beam. In a
clumpy medium the escaping line radiation would then
consist of an appropriate angle-averaged spectrum over
a distribution of such ``blobs'' (e.g. see also Nandra \& George
1994; Miller, Turner, \& Reeves, 2009 for discussion of such scenarios).
The appropriate mean column density, \meannhp, for an externally-illuminated
sphere is $2/3$ of the equatorial column density.
We find that for an individual blob, not surprisingly,
the maximum
\fekalfa line efficiency, $\sim 2.6\%$, is achieved for line photons
escaping in a
direction anti-parallel to
the incident radiation, from the illuminated face, for \meannh~$\sim 
8 \times 10^{23} \rm \ cm^{-2}$. For this column density,
the maximum efficiency for \fekalfa line emission
from the opposite face, parallel to the
incident radiation is $2.1\%$, for 
\meannh~$\sim  
6$--$7 \times 10^{23} \rm \ cm^{-2}$. The angle-averaged
value of \fekaeffic (over $0$--$\pi$) peaks at $2.2\%$
for \meannh~$\sim
5$--$6 \times 10^{23} \rm \ cm^{-2}$
In addition to the angular averaging, for a clumpy medium
one must account for the volume filling factor
and covering factor of such blobs, and for interaction
of line photons with multiple clouds. These effects
can decrease the maximum \fekalfa line efficiency, and in fact,
as the covering factor and filling factor tend towards unity,
\fekaeffic has to tend to the value for a centrally-illuminated
sphere, or $\sim 1.6\%$.
For any reprocessor geometry, the general behavior
of \fekaeffic as a function of mean column density is
the same. That is, a rise to a maximum, followed by a decline
that is steeper the more heavily obscured
the line of sight is. The column density for which the maximum
is achieved will be in the range $\sim 3-8 \times
10^{23} \ \rm cm^{-2}$, with the largest values of
\meannh and \fekaeffic at the maximum
corresponding to unobscured (pure reflection) lines of sight.
However, none
of the geometries that we have considered here 
(for cosmic abundances and neutral matter)
give a value of \fekaeffic greater than $\sim 3\%$ (for $\Gamma=1.9$),
for any covering factor, column density, or line of sight.

Our results also emphasize that the \fekalfa emission line
cannot be trivially used as a proxy for the intrinsic 
continuum luminosity in obscured AGN. It has been postulated
that since the \fekalfa line luminosity samples the
intrinsic continuum over a large fraction of the sky,
as seen from the X-ray source, it might be a good indicator
of the intrinsic continuum luminosity, given that the
line-of-sight may not be representative of the column
density for the bulk of the reprocessor. However, we have
shown that the dependence of the luminosity of the
\fekalfa line on the geometry, inclination angle, \nhp, and
\afe is sufficiently complex that such a use of the \fekalfa
line cannot be trivially implemented. In particular, as
a function of \nhp, the \fekalfa line luminosity is 
double-valued. However, we can obtain a lower limit on
the intrinsic luminosity.
We have shown that regardless of geometry, inclination,
and the column density of the reprocessor, for a given
$\Gamma$ and \afep, there is a maximum possible value for
\fekaeffic if one can be certain that
the dominant species of Fe are Fe~{\sc xii} or less.
We can measure $\Gamma$, and there may be
independent constraints on \afep. Therefore, if we measure
a flux or luminosity of the \fekalfa line,
the lower bound on such
a measurement implies a lower limit on the intrinsic 
continuum luminosity that corresponds to the maximum efficiency.
  
\section{Application to absorption-dominated models for MCG~$-$6-30-15}
\label{application}

We can apply our results to the case of
the well-known Seyfert galaxy MCG~$-$6-30-15. This AGN has
been the archetypal X-ray source for harboring
a broad, relativistic \fekalfa line. However,
Miller, Turner, \& Reeves (2008) showed that 
absorption-dominated models that do not require a
relativistically-broadened \fekalfa line
are also consistent with the X-ray data.
Reynolds \etal (2009) claimed
that such absorption-dominated models over-predict
the flux of the narrow \fekalfa emission-line at $\sim 6.4$~keV
and concluded 
that a finely-tuned model with a small covering factor is required.
Here we show that this conclusion
is invalid on the basis of simple physics.

MCG~$-$6-30-15 has a typical narrow \fekalfa line at 6.4~keV
with a measured flux of $1.6^{+1.1}_{-0.9} \times 10^{-5}
\ \rm photons \ cm^{-2} \ s^{-1}$ (Yaqoob \& Padmanabhan 2004).
The time-averaged 2--10 keV flux from
a very long $\sim 300$~ks \xmm observation 
(Fabian \etal 2002; Dov\v{c}iak \etal 2004) was $\sim 4.3
\times 10^{-11} \rm \ erg \ cm^{-2} \ s^{-1}$ (a value
this is representative of historical behavior--see
Markowitz \etal 2003), and $\Gamma=1.9$ is typical. 
Therefore, we obtain a photon flux above the Fe~K edge of
$2.8   \times 10^{-3} \ \rm photons \ cm^{-2} \ s^{-1}$.
Using our general result of a maximum 
\fekalfa line efficiency of $\sim 3\%$ we get
an \fekalfa line flux of
$8.5 \times 10^{-5} \ \rm photons \ cm^{-2} \ s^{-1}$.
Thus, no matter what distribution of cold/neutral matter
that we place around the X-ray source in MCG~$-$6-30-15,
regardless of column density, covering factor, or geometry,
we cannot obtain an \fekalfa line flux much greater than
$8.5 \times 10^{-5} \ \rm photons \ cm^{-2} \ s^{-1}$.
Miller \etal (2009) presented similar conclusions for the
narrow \fekalfa line flux in MCG~$-$6-30-15.
This is a very generous upper limit because 
it requires optimal values for the column density
and covering factor, and a line-of-sight that does not intercept
the line emitter.
Obviously, a large column density in the line-of-sight will
reduce the predicted \fekalfa line flux. For example,
for $N_{H}=2 \times 10^{24} \ \rm cm^{-2}$, and a spherical
geometry with a covering factor of 0.5, the predicted
\fekalfa line flux is 
very small, just $\sim 4 \times 10^{-6} \ \rm photons \ cm^{-2} \ s^{-1}$. 
A similar value is obtained from the MY09 edge-on toroidal
X-ray reprocessor model (see \figefficvsnhp).
Reynolds \etal (2009) obtained estimates of the \fekalfa line flux
in MCG~$-$6-30-15 based on a simple cold,
cosmic abundance absorber with $N_{H} = 2 \times 10^{24} \ \rm cm^{-2}$.
However, they used an optically-thin approximation for
a manifestly optically-thick scenario. Specifically, they used
a column density that is one hundred times higher than the
value for which that approximation
begins to breaks down (see \figefficvsnhp), and consequently obtained
an effective efficiency that is unphysical. 
Reynolds \etal (2009) therefore 
obtained a very high \fekalfa line flux ($2.54 \times 10^{-4}
\rm \ photons \ cm^{-2} \ s^{-1}$) that is nearly two
orders of magnitude higher than that expected for their
assumed column density and covering factor 
($N_{H}=2 \times 10^{24} \ \rm cm^{-2}$ and 0.35 respectively).
They also gained a factor of $\sim 2$ in the \fekalfa line flux
by assuming that {\it all}
of the continuum photons absorbed above the Fe~K edge are
absorbed by the Fe~K shell, whereas in fact the true fraction is 0.51--0.61
(see \S\ref{lineflux}).

\section{Summary}
\label{torusconcl}

The \fekalfa emission-line flux or luminosity is nontrivial
to calculate because the optically-thin approximation 
(with respect to absorption and scattering in the
$\sim 6$--$7$~keV band) breaks
down for column densities as small as $\sim 2 \times 10^{22}
\ \rm cm^{-2}$. Inappropriate use of the optically-thin approximation
can lead to estimates of the line flux that are orders of
magnitude too large. 
We have given a prescription for estimating the \fekalfa line
flux in the case of neutral
matter with cosmic abundances, for some very general situations. 

We have
characterized the line flux
in terms of a dimensionless {\it efficiency}, \fekaefficp,
defined as the ratio
of the flux in the line to that in the incident continuum
above the Fe~K absorption edge threshold energy. 
We refer to values of \fekaeffic that pertain to a
power-law slope of the incident X-ray continuum of $\Gamma=1.9$
but have given a simple prescription for scaling
\fekaeffic to other values of $\Gamma$ in the range 1.5--2.5. 
The maximum value of \fekaefficp depends on the
geometry, covering factor, and orientation, and is always obtained
for lines of sight to the X-ray reprocessor that are
unobscured. In general, this maximum is achieved before
the structure becomes Compton-thick, for mean column densities
in the range $\sim 3-8 \times 10^{23} \rm \ cm^{-2}$.
For the toroidal geometry of MY09 
the maximum \fekalfa line efficiency is attained for nearly full
covering ($[\Delta\Omega/(4\pi)] \sim 0.9$),
before the medium
becomes Compton-thick, and is $\sim 2.8\%$. 
This can be compared with the maximum value of
\fekaeffic for a uniform, centrally-illuminated sphere of $\sim 1.6\%$.
A clumpy distribution of matter can give somewhat higher \fekalfa line
efficiencies, of the order of $2\%$.

Regardless of geometry and covering factor,
for cosmic abundances and
incident continua with a power-law photon index of 1.9 or
less, none of the geometries considered yields
an \fekalfa line efficiency greater than
$\sim 3\%$. 
Steeper incident continua can give a higher efficiency but 
we find that for $\Gamma=2.5$ it is only 
a factor of $\sim 1.43$ higher.
X-ray sources that are Compton-thick in the
line of sight can have \fekalfa line efficiencies that are
an order of magnitude or more smaller than the maximum for
a given geometry. For example the MY09 toroidal geometry viewed
edge-on has \fekaefficp~$< 0.2\%$, and \fekaefficp~$< 0.02\%$
for column densities of 
$2 \times 10^{24} \ \rm cm^{-2}$ and $10^{25} \ \rm cm^{-2}$
respectively.
We have applied our results to the Seyfert galaxy MCG~$-$6-30-15
to show that absorption-dominated models for the broad,
relativistic \fekalfa emission line {\it do not} over-predict
the flux of the narrow \fekalfa line 
for any column density or covering factor. Recent claims to the
contrary by Reynolds \etal (2009) are based on an 
invalid treatment of the problem.
Our results also emphasize the fact that the absolute luminosity
of the \fekalfa emission line in obscured astrophysical sources
cannot be used as a proxy for the intrinsic continuum luminosity
in a trivial way. Using both the \fekalfa line flux {\it and}
EW does provide additional constraints
but the intrinsic continuum luminosity may still
be uncertain because of the 
complex dependencies upon
Fe abundance, geometry, covering factor, and system orientation. 

The authors thank Chris Done for carefully reviewing the
manuscript and for helpful suggestions. 
Partial support from NASA grants NNG04GB78A (KM, TY) and
NNX09AD01G (TY) is gratefully acknowledged. TJT acknowledges
support from NASA grant NNX08AJ41G.

\end{document}